# A Mathematical Model of Transmission Dynamics of SARS-Cov-2 (Covid-19) with an Underlying Condition of Diabetes

Samuel Okyere[*1], Joseph Ackora-Prah[1], Ebenezer Bonyah[2]

[1]Department of Mathematics, Kwame Nkrumah University of Science and Technology, Kumasi, Ghana

[2]Department of Mathematics Education, Akenten Appiah-Menka University of Skills Training and Enterpreneurial Development, Kumasi, Ghana

Corresponding Author's Email: okyere2015@gmail.com

**Abstract**

It is well established that people with diabetes are more likely to have serious complications from COVID-19. Nearly 1 in 5 COVID-19 deaths in the African region are linked to diabetes. World Health Organization (WHO) finds that 18.3% of COVID-19 deaths in Africa are among people with diabetes. In this paper, we have formulated and analysed a mathematical comorbidity model of diabetes - COVID-19 of the deterministic type. The basic properties of the model were explored. The basic reproductive number, equilibrium points and stability of the equilibrium points were examined. Sensitivity analysis of the model was carried on to determine the impact of the model parameters on the basic reproductive number $(R_0)$ of the model. The model had a unique endemic equilibrium point, which was stable for $R_0 > 1$. Time-dependent optimal controls were incorporated into the model with the sole aim of determining the best strategy for curtailing the spread of the disease. COVID-19 cases from March to September 2020 in Ghana were used to validate the model. Results of the numerical simulation suggest a greater number of individuals deceased when the infected individual had an underlying condition of diabetes. More so COVID-19 is endemic in Ghana with the basic reproduction number found to be $R_0 = 1.4722$. The numerical simulation of the optimal control model reveals the lockdown control minimized the rate of decay of the susceptible individuals whereas the vaccination led to a number of susceptible individuals becoming immune to COVID-19 infections. In all the two preventive control measures were both effective in curbing the spread of the COVID-19 disease as the number of COVID-19 infections was greatly reduced. We conclude that more attention should be paid to COVID-19 patients with an underlying condition of diabetes as the probability of death in this population was significantly higher.

# 1 Introduction

Diabetes is one of the underlying conditions associated with a high risk of COVID-19 complications. World Health Organization finds that 18.3% of COVID-19 deaths in Africa are among people with diabetes [9]. A recent WHO analysis evaluated data from 13 countries on underlying conditions or comorbidities in Africans who tested positive for COVID-19 revealed a 10.2% case fatality rate in patients with diabetes, compared with 2.5% for COVID-19 patients overall [10].

Having heart failure, coronary artery disease, and hypertension can make you more severely ill from COVID-19. However, the case fatality rate for people with diabetes is twice as high as the fatality rate among patients suffering from any comorbidity mentioned [10]. Studies have also shown that COVID-19 does not affect all population groups equally. The risk of severe COVID-19 increases as the number of underlying medical conditions increases in a person [5-8].

Research conducted by Choi (2021) [32], on the mortality rates of 566,602 patients with coronavirus disease (COVID-19) in a Mexican community revealed the mortality rate of patients with the underlying health conditions was 12% and was four (4) times higher than that of patients without the underlying condition. Several studies have confirmed that COVID-19 is more severe in older people and those with the underlying condition of diabetes, lung or heart diseases [11].

Statistics from the World Health Organization (WHO) indicates that globally over 460 million cases of COVID-19 have been recorded with more than 6 million deaths [12]. According to WHO, 422 million people worldwide have diabetes, particularly in low-and middle-income countries and 1.5 million deaths are directly attributed to diabetes each year [13]. In Ghana, studies on the general population have estimated that between 3.3 and 6% of the population has diabetes with the prevalence increasing with age and being higher in urban than in rural areas [14]. Research conducted by Oduro-Mensah et al., (2020) [15], reveals diabetes (9.9%) as the second-highest underlying condition of COVID-19 infected patients in two national treatment centres in Ghana which were GA East Municipal hospital and the University of Ghana Medical Center.

Research on COVID-19 has been conducted from various perspectives such as infection cases, clinical characteristics and preventive measures [19, 33]. The transmission dynamics of infectious diseases have been studied and analyzed by researchers using mathematical models [17-26]. Mathematical models play an important role in understanding the complexities of infectious diseases and their control [29]. Existing studies on COVID-19 using mathematical models have targeted the transmission dynamics in a population without taking into consideration the underlying health conditions of the patients [17-24]. In [22], the authors formulated a two-patch mathematical model with a mobility matrix to capture the spatial heterogeneity of COVID-19 outbreaks in South Korea. In [18], the authors formulated a mathematical model that incorporates the currently known disease characteristics and tracks various intervention measures in Uganda. In [21], the authors proposed an age-structured Susceptible-Latent-Mild-Critical-Removed (SLMCR) compartmental model to study the transmission rate of COVID-19 disease in six countries which were Australia, Italy, Spain, the USA, the UK, and Canada.

Some authors have also proposed models to examine COVID-19 mortality rates in various countries [37-39]. In [37], the authors proposed an exponentiated transformation of Gumbel Type-II (ETGT-II) for modeling the two data sets of death cases due to COVID-19 in Europe and China.

Few papers have considered a population with comorbidity(s). In [17], the authors proposed a deterministic-epidemic model that describes the spreading of SARS-COV-2 within a community with comorbidities incorporating a size-dependent area to quantify the effect of social distancing. In [31], the authors proposed a mathematical model to investigate the transmission dynamics of COVID-19 with an emphasis on the relationship between the disease transmission and the chronic health conditions of the host population. The authors in [31] partitioned the population into two separate groups, one with an underlying condition and the other without an underlying condition and described the disease transmission within and among the groups.

The current study is aimed to propose a new compartmental model using differential equations to describe the transmission of COVID-19 in a population with an underlying condition of diabetes. Here the model solely focuses on human-to-human transmission of COVID-19 in a population with a portion having an underlying condition of diabetes. The study is motivated by the available COVID-19 works which has been reviewed. Comorbidity of diabetes – COVID -19 has not yet been investigated according to literature.

The subsequent sections of the paper are as follows: In Sect. 2, we propose and explore a mathematical model of COVID-19 with comorbidity of diabetes. In sect. 3, we determine the qualitative features of the model. Sect. 4, analysis of the model formulated. In sect. 5, we incorporate the optimal control in the formulated model. In sect. 6, we analyzed numerically the behaviour of the formulated model using available data. Finally, in Sect. 6, we discuss and conclude the results of our proposed study.

## 2  Model Formulation

We examine the transmission dynamics of the diabetes-COVID-19 through modifying the model of [31] by introducing a population with diabetes and establishing the transmission within the population. With an idea of the transmission dynamics of COVID-19 and the incidence of diabetes [19, 20], we formulate a new model to describe the interactions of a population with diabetes in the event of COVID-19 transmission. We classify the total population $N(t)$ into six groups: Susceptible individuals $S(t)$; susceptible with diabetes $D(t)$; individuals exposed to COVID-19 $E(t)$; COVID-19 infected individuals $I(t)$; COVID-19 patients with diabetes $C(t)$; and individuals removed from COVID-19 $R(t)$.

We assume that the population is homogeneously mixed, with no restriction of age, mobility or other social factors. All newborns are susceptible (no inherited immunity). Individuals are recruited into the susceptible population at the rate $\Omega$ and they become diabetic at the rate $\lambda$. The susceptible may be infected when they interact with those in class I or C. Then the infected person becomes exposed to the disease and hence join the class E at the rate $\beta$. The proportion of the exposed that joined the compartment I and C are $\alpha \varphi E$ and $(1-\alpha)\varphi$ respectively. The parameter $\mu, \delta, \delta_1$ and $\delta_2$ are the natural mortality rate, the disease-induced death rate of COVID-19 infected people, COVID-19 patients with diabetes and susceptible with diabetes respectively. The rate at which COVID-19 patients with diabetes recover from COVID-19 is $\gamma_1$ and the rate at which COVID-19 infected people recover is $\gamma$. The flowchart of the model is shown in Fig. 1.

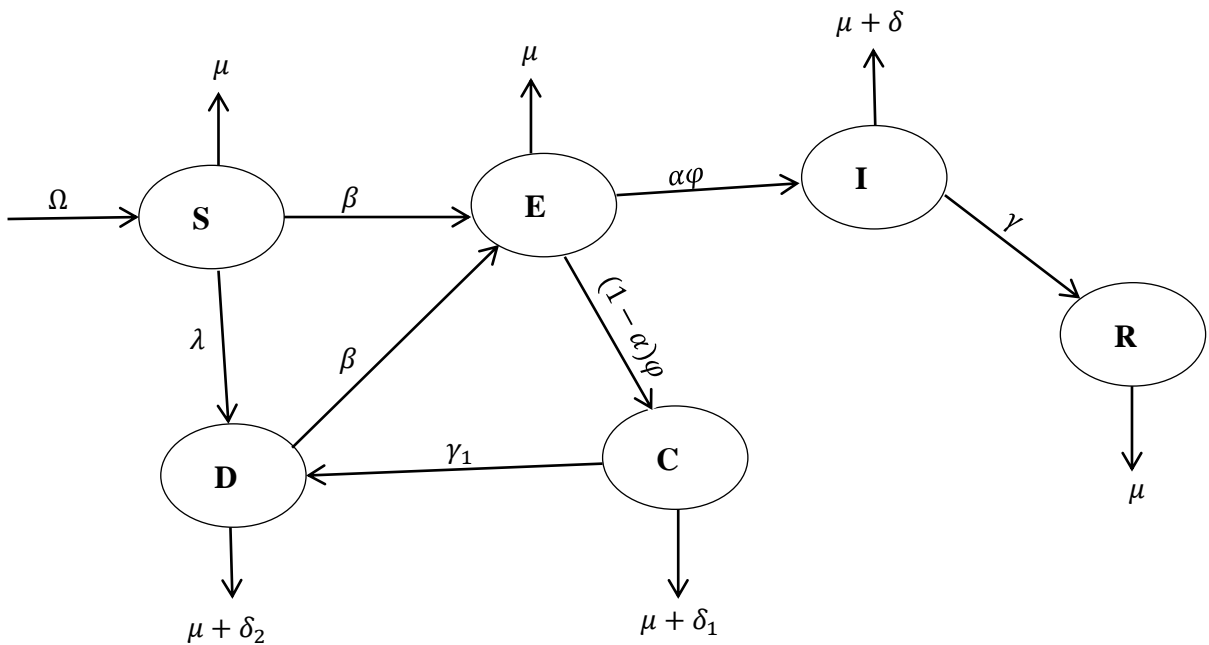

Fig.1: Schematic diagram of diabetes – COVID-19 Model

The following ordinary differential equations describe the model:

$$\frac{dS}{dt} = \Omega - \beta\left(\frac{I+C}{N}\right)S - (\mu + \lambda)S,$$

$$\frac{dD}{dt} = \lambda S + \gamma_1 C - \beta\left(\frac{I+C}{N}\right)D - (\delta_2 + \mu)D,$$

$$\frac{dE}{dt} = \beta\left(\frac{I+C}{N}\right)S + \beta\left(\frac{I+C}{N}\right)D - (\varphi + \mu)E, \quad (1)$$

$$\frac{dI}{dt} = \alpha\varphi E - (\delta + \mu + \gamma)I,$$

$$\frac{dC}{dt} = (1-\alpha)\varphi E - (\delta_1 + \mu)C,$$

$$\frac{dR}{dt} = \gamma I - \mu R.$$

With initial conditions $S(0) = S_0, D(0) = D_0, E(0) = E_0, I(0) = I_0, C(0) = C_0, R(0) = R_0$.

## 3 Qualitative Analysis of the Proposed Model

This section presents the computation of the invariant region and the basic reproduction number for the proposed model (1) and study the locally asymptotically stability of the equilibrium points

### 3.1 Invariant Region

We present the following results which guarantee that system (1) is epidemiologically and mathematically well-posed in a feasible region $\Phi$, given as

$$\Phi = \left[(S, D, E, I, C, R) \in R_+^6 : N \leq \frac{\Omega}{\mu}\right] \quad (2)$$

**Theorem 3.1**: There exists a domain $\Phi$ in which the solution set $(S, D, E, I, C, R)$ is contained and bounded.

Proof:

Given the solution set $(S, D, E, I, C, R)$ with positive initial conditions

$$S(0) = S_0, D(0) = D_0, E(0) = E_0, I(0) = I_0, C(0) = C_0, R(0) = R_0$$

We let, $N(t) = S(t) + D(t) + E(t) + I(t) + C(t) + R(t)$, then

$$N'(t) = S'(t) + D'(t) + E'(t) + I'(t) + C'(t) + R'(t)$$

It follows that $N'(t) \prec \Omega - \mu N$

Solving the differential inequalities yields

$$N'(t) \leq \frac{\Omega}{\mu} + N(0)e^{-\mu(t)}$$

Taking the limits as $t \to \infty$, gives $N' \leq \frac{\Omega}{\mu}$

That is, all solutions are confined in the feasible region $\Phi$. We now show that the solutions of system (1) are nonnegative in $\Phi$.

**Theorem 3.2:** Having describe the human population in the model (1), it is vital to show that the state parameters $S(t), D(t), E(t), I(t), C(t), R(t)$ are nonnegative for all $t > 0$ in the domain $\Phi$.

*Proof:*

It is easy to see $S(t) > 0$, for all $t \geq 0$. If not, let there exist $t_* > 0$ such that $S(t_*) = 0, S'(t_*) \leq 0$ for all $0 \leq t \leq t_*$. Then, from the first equation of the model (1), we have

$$\frac{d}{dt}\left(Se^{\left(\frac{\beta I+\beta C}{N}+\mu+\lambda\right)t}\right)=\Omega e^{\left(\frac{\beta I+\beta C}{N}+\mu+\lambda\right)t}$$

Integrating from 0 to $t*$ we obtain

$$S(t_*)e^{\left(\frac{\beta I+\beta C}{N}+\mu+\lambda\right)t} - S(0) = \int_0^{t_*}\Omega e^{\left(\frac{\beta I+\beta C}{N}+\mu+\lambda\right)\tau}d\tau$$

Multiplying through by $e^{\left(\frac{\beta I+\beta C}{N}+\mu+\lambda\right)t}$, we obtain

$$S(t_*) = S(0)e^{-\left(\frac{\beta I+\beta C}{N}+\mu+\lambda\right)t} + e^{-\left(\frac{\beta I+\beta C}{N}+\mu+\lambda\right)t}\left[\int_0^{t_*}\Omega e^{\left(\frac{\beta I+\beta C}{N}+\mu+\lambda\right)\tau}d\tau\right] > 0$$

which contradicts $S(t_*) = 0$.

Similarly, from the remaining five (5) equations of system (1), the following results can be obtained

$$D(t_*) = D(0)e^{-\left(\frac{\beta I+\beta C}{N}+\delta_2+\mu\right)t} + e^{-\left(\frac{\beta I+\beta C}{N}+\delta_2+\mu\right)t}\left[\int_0^{t_*}[\lambda S + \gamma_1 C]e^{\left(\frac{\beta I+\beta C}{N}+\delta_2+\mu\right)\tau}d\tau\right] > 0$$

$$E(t_*) = E(0)e^{-(\varphi+\mu)t} + e^{-(\varphi+\mu)t}\left[\int_0^{t_*}\beta\left(\frac{(I(\tau)+C(\tau))S(\tau)+(I(\tau)+C(\tau))D(\tau)}{N(\tau)}\right)e^{(\varphi+\mu)\tau}d\tau\right] > 0,$$

$$I(t_*) = I(0)e^{-(\delta+\mu+\gamma)t} + e^{-(\delta+\mu+\gamma)t}\left[\int_0^{t_*}\alpha\varphi E(\tau)e^{(\delta+\mu+\gamma)\tau}d\tau\right] > 0,$$

$$C(t_*) = C(0)e^{-(\delta_1+\mu)t} + e^{-(\delta_1+\mu)t}\left[\int_0^{t_*}(1-\alpha)\varphi E(\tau)e^{(\delta_1+\mu)\tau}d\tau\right] > 0,$$

$$R(t_*) = R(0)e^{-\mu t} + e^{-\mu t}\left[\int_0^{t_*} \gamma I(\tau)e^{\mu\tau}d\tau\right] > 0.$$

which contradicts $D(t_*) = E(t_*) = I(t_*) = C(t_*) = R(t_*) = 0$. Hence this completes the proof.

## 3.2 Stability of the Equilibrium Points

The disease–free equilibrium $(E_0)$ is the steady-state solution where there is no COVID-19 infection in the population. Setting $E = I = C = 0$ and the right-hand side of the system (1) to zero, then solving yields

$$E_0 = (S^0, D^0, E^0, I^0, C^0, R^0) = \left(\frac{\Omega}{\mu + \lambda}, \frac{\Omega\lambda}{(\delta_2 + \mu)(\mu + \lambda)}, 0,0,0,0\right) \quad (3)$$

We denote the endemic equilibrium point $(E_1)$ by $E_1 = (S^*, D^*, E^*, I^*, C^*, R^*)$. Equating the right-side of the system (1) to zero and solving yields

$$\begin{cases} S^* = \dfrac{\Omega}{\beta\dfrac{(I^* + C^*)}{N} + (\mu + \lambda)}, \quad D^* = \dfrac{\lambda S^* + \gamma_1 C^*}{\beta\dfrac{(I^* + C^*)}{N} + (\mu + \delta_2)} \\ E^* = \dfrac{\beta}{(\varphi + \mu)}\left[\dfrac{(I^* + C^*)(S^* + D^*)}{N}\right], \quad I^* = \dfrac{\alpha\varphi E^*}{(\delta + \mu + \gamma)} \\ C^* = \dfrac{(1-\alpha)\varphi E^*}{(\delta_1 + \mu + \gamma_1)}, \quad R^* = \dfrac{\gamma I^*}{\mu} \end{cases} \quad (4)$$

We now determine the basic reproduction number of the model (1). The basic reproduction number is the number of secondary cases produced in a susceptible population by a single infective individual during the time of the infection. We evaluate the basic reproduction number using the next generation operator method [3]. From the model (1), $E, I$ and $C$ are the COVID-19 infected compartments. We decomposed the right-hand side of system (1) corresponding to the COVID-19 infected compartments as $F - V$, where

$$F = \begin{pmatrix} \beta\left(\dfrac{I+C}{N}\right)S + \beta\left(\dfrac{I+C}{N}\right)D \\ \alpha\varphi E \\ (1-\alpha)\varphi E \end{pmatrix} \text{ and } V = \begin{pmatrix} (\varphi + \mu)E \\ (\delta + \mu + \gamma)I \\ (\delta_1 + \mu)C \end{pmatrix}.$$

Next, we find the derivative of $F$ and $V$ evaluated at the disease-free steady state and this gives the matrices

$$F = \dfrac{\partial F}{\partial x_i} = \begin{pmatrix} 0 & \dfrac{\beta(S^0 + D^0)}{N} & \dfrac{\beta(S^0 + D^0)}{N} \\ \alpha\varphi & 0 & 0 \\ (1-\alpha)\varphi & 0 & 0 \end{pmatrix} \text{ and } V = \dfrac{\partial V}{\partial x_i} = \begin{pmatrix} \varphi + \mu & 0 & 0 \\ 0 & \delta + \mu + \gamma & 0 \\ 0 & 0 & \delta_1 + \mu \end{pmatrix}.$$

Where $x_i = E, I, C$

$$FV^{-1} = \begin{pmatrix} 0 & (\delta + \mu + \gamma)\beta\left(\dfrac{S^0 + D^0}{N}\right) & 0 \\ \alpha\varphi(\varphi + \mu) & 0 & 0 \\ (1-\alpha)\varphi & 0 & 0 \end{pmatrix}.$$

The basic reproduction number is the largest positive eigenvalue of $FV^{-1}$ and is given as

$$R_0 = \sqrt{\beta\alpha\varphi(\varphi + \mu)(\delta + \mu + \gamma)\left(\dfrac{\Omega(\lambda + \delta_2 + \mu)}{(\mu + \lambda)(\delta_2 + \mu)}\right) + (1-\alpha)\varphi}. \qquad (5)$$

## 3.3 Local Stability of the Disease-free Equilibrium Point

The necessary condition for the local stability of the disease –free steady state is established in the following theorem.

Theorem 3.3: The disease-free equilibrium (DFE) is locally asymptotically stable if $R_o < 1$ and unstable for $R_o > 1$.

**Proof:**

The Jacobian matrix with respect to the system (1) is given by,

$$J = \begin{bmatrix} J_{11} & 0 & 0 & -\dfrac{\beta S}{N} & -\dfrac{\beta S}{N} & 0 \\ \lambda & J_{22} & 0 & -\dfrac{\beta D}{N} & -\dfrac{\beta D}{N} & 0 \\ \beta\left(\dfrac{I+C}{N}\right) & \beta\left(\dfrac{I+C}{N}\right) & -(\varphi+\mu) & \dfrac{\beta S}{N}+\dfrac{\beta D}{N} & \dfrac{\beta S}{N}+\dfrac{\beta D}{N} & 0 \\ 0 & 0 & \alpha\varphi & -(\gamma+\mu+\delta) & 0 & 0 \\ 0 & 0 & (1-\alpha)\varphi & 0 & -(\delta_1+\mu) & 0 \\ 0 & 0 & 0 & \gamma & 0 & -\mu \end{bmatrix}. \quad (6)$$

Where,

$$J_{11} = -\beta\left(\dfrac{I+C}{N}\right) - (\lambda+\mu),\ J_{22} = -\beta\left(\dfrac{I+C}{N}\right) - (\delta_2+\mu).$$

The Jacobian matrix evaluated at the disease-free equilibrium point is given as

$$J_{E^0} = \begin{bmatrix} -(\lambda+\mu) & 0 & 0 & -\dfrac{\beta S^0}{N} & -\dfrac{\beta S^0}{N} & 0 \\ \lambda & -(\delta_2+\mu) & 0 & -\dfrac{\beta D^0}{N} & -\dfrac{\beta D^0}{N} & 0 \\ 0 & 0 & -(\varphi+\mu) & \dfrac{\beta S^0}{N}+\dfrac{\beta D^0}{N} & \dfrac{\beta S^0}{N}+\dfrac{\beta D^0}{N} & 0 \\ 0 & 0 & \alpha\varphi & -(\gamma+\mu+\delta) & 0 & 0 \\ 0 & 0 & (1-\alpha)\varphi & 0 & -(\delta_1+\mu) & 0 \\ 0 & 0 & 0 & \gamma & 0 & -\mu \end{bmatrix}. \quad (7)$$

From the system (7), the first three (3) eigenvalues are $-(\lambda+\mu), -(\delta_2+\mu)$ and $-\mu$. The rest are obtained by deleting the first, second, and sixth rows and columns of the system (7). This gives

$$J_{E^0} = \begin{bmatrix} -\varphi-\mu & A & A \\ \alpha\varphi & -(\delta+\mu+\gamma) & 0 \\ (1-\alpha)\varphi & 0 & -(\delta_1+\mu) \end{bmatrix}. \quad (8)$$

Where

$$A = \beta\left(\dfrac{S^0+D^0}{N}\right).$$

The characteristic equation of the system (8) is

$$\Phi(\omega) = \omega^3 + A_1\omega^2 + A_2\omega + A_3 = 0, \quad (9)$$

where

$$A_1 = \delta^\psi + 3\mu^\psi + \varphi^\psi + \gamma^\psi + \delta_1^\psi,$$

$$A_2 = \delta_1^\psi(\delta^\psi + \gamma^\psi + \varphi^\psi + 2\mu^\psi) + \mu^\psi(3\mu^\psi + 2(\delta^\psi + \varphi^\psi + \gamma^\psi)) + \varphi^\psi(\delta^\psi + \gamma^\psi) + \frac{(1-\alpha)\varphi^\psi - R_0^2}{\alpha(\delta^\psi + \mu^\psi + \gamma^\psi)},$$

$$A_3 = \mu^{2\psi}(\alpha + \varphi^\psi + \delta_1^\psi + \gamma^\psi + \mu) + (\alpha\delta_1^\psi + \delta^\psi + \gamma^\psi - \alpha(\delta^\psi + \gamma^\psi))\left[\frac{(1-\alpha)\varphi^\psi - R_0^2}{\alpha(\delta^\psi + \mu^\psi + \gamma^\psi)}\right].$$

From Routh – Hurwitz stability criterion [23], if the conditions $A_1 > 0, A_2 > 0, A_3 > 0$ and $A_1 A_2 - A_3 > 0$ are true, then all the roots of the characteristic equation (10) have a negative real part which means stable equilibrium. The coefficients $A_1 \geq 0, A_2 \geq 0, A_3 \geq 0$ and the condition

$$A_1 A_2 - A_3 = \varphi^\psi(\delta^\psi + \gamma^\psi) + \delta_1^\psi(\varphi^\psi + \delta^\psi + \gamma^\psi + 2\mu^\psi) + 2\mu^\psi(\varphi^\psi + \delta^\psi + \gamma^\psi + 2\mu^\psi) - \mu^{2\psi}(\alpha^\psi + \delta_1^\psi$$
$$+ \varphi^\psi + \gamma^\psi + \mu^\psi) + (\varphi^\psi + 2\delta^\psi + 2\gamma^\psi + 3\mu^\psi - \alpha^\psi\delta_1^\psi - \alpha^\psi(\delta^\psi + \gamma^\psi))\left[\frac{(1-\alpha)\varphi^\psi - R_0^2}{\alpha(\delta^\psi + \mu^\psi + \gamma^\psi)}\right] > 0$$

when $R_0 < 1$. In the case when $R_0 > 1$, we have that at least $A_2 < 0$ and by using Decartes rule of sign, we can conclude that one of the eigenvalues is positive. Therefore the system is unstable. This completes the proof.

## 4 Numerical Analysis of the Diabetes-COVID-19 Model

In this section, the model is validated using confirmed cases data from Ghana Health Service for the period March -- September 2020 [26]. Using Matlab Gaussfit, the cumulative data of confirmed COVID-19 cases for the period March -- September 2020 with the best fitted curve is depicted in Fig. 2. Fig. 3 shows the residuals of the best fitted curve.

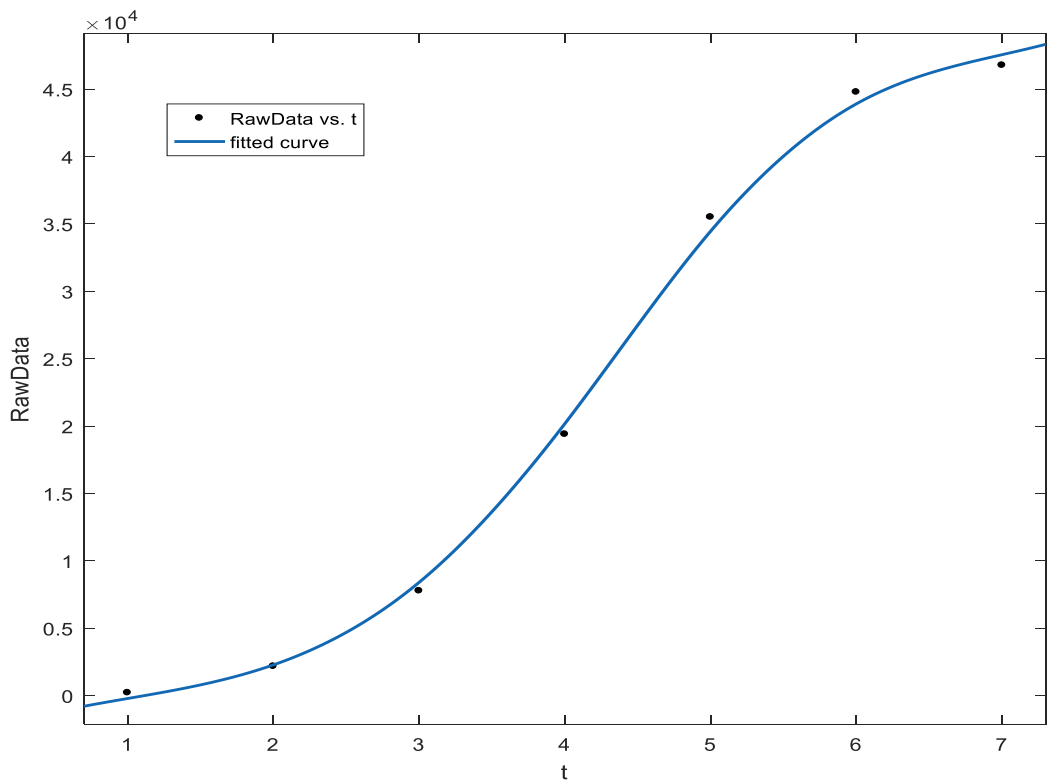

Fig.2: Cumulative cases of Ghana's COVID-19 from March - September 2020 with the best-fitted curve.

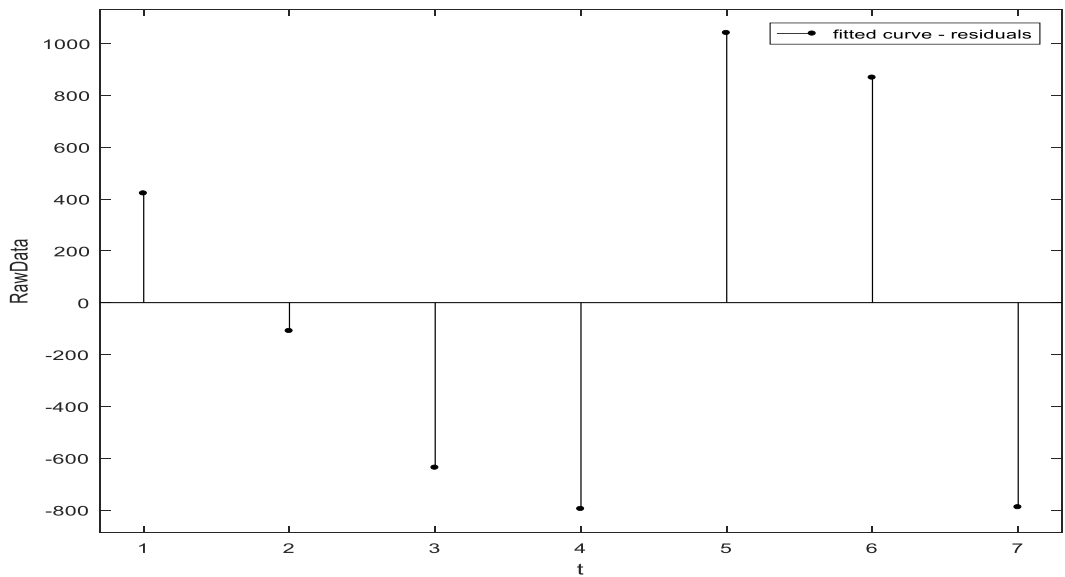

Fig. 3 Residuals of the best-fitted curve

Table 1: Parameter values and description

| Parameter | Description | Value | Source |
|---|---|---|---|
| $\Omega$ | The recruitment rate | 28.452 | [2] |
| $\beta$ | The transmission rate | 0.9 | Estimated |
| $\delta$ | COVID-19 disease-induced death | 0.0016728 | [1] |
| $\lambda$ | Incidence of diabetes | 0.2 | [20] |
| $\gamma$ | The rate of recovery of COVID-19 patients with no underlying condition of diabetes | 1/14 | Estimated |
| $\mu$ | The natural mortality rate | $0.4252912 \times 10^{-4}$ | Estimated |
| $\varphi$ | The rate at which the exposed become infectious | 0.25 | [16] |
| $\gamma_1$ | The recovery rate of COVID-19 patients with an underlying condition of diabetes | 1/14 | Estimated |
| $\delta_1$ | Disease-induced death rate of COVID-19 patients with an underlying condition | 0.0144 | [31, 34] |
| $\delta_2$ | Disease-induced death rate of diabetes | 0.05 | [20] |

For the parameters used in the simulation (see Table 1), one computes the basic reproductive number and obtain $R_0 = 1.4722$ which shows the disease is endemic in Ghana. Next, we investigate the sensitiveness of the model (1) with respect to the variation of each one of its parameters for the endemic threshold (5).

## 5.1 Sensitivity Analysis

Sensitivity analysis tells us how important each parameter is to the disease transmission. We use the normalized forward sensitivity index of a variable with respect to a given parameter of the model (1). This is given as

$$\chi_\theta^{R_0} = \frac{\partial R_0}{\partial \theta} \times \frac{\theta}{R_0}, \tag{10}$$

where $\theta$ is the parameter under consideration. A positive sensitivity index means an increase in the value of the parameter would lead to a percentage increase in the basic reproduction number and a decrease in the parameter would decrease the basic reproductive number. Using the parameter values in Table 1 and the threshold parameter (5), we determine the sensitivity indices and is given in Table 2.

Table 2: Sensitivity Indices of the Model Parameters

| Parameter | Value |
|---|---|
| $\Omega$ | 0.4827 |
| $\beta$ | 0.1973 |
| $\delta$ | 0.00037962 |
| $\lambda$ | $-2.6134 \times 10^3$ |
| $\gamma$ | 0.0162 |
| $\mu$ | $3.6233 \times 10^{-4}$ |
| $\varphi$ | 2.9130 |
| $\gamma_1$ | 0.0 |
| $\delta_1$ | 0.0 |
| $\delta_2$ | −0.4036 |

The most sensitive parameters to the basic reproduction number are $\varphi, \Omega, \beta, \delta_2$ and $\lambda$. In concrete, an increase of the value of $\beta$ will increase the basic reproduction number by 19.73%. In contrast, an increase of the value of $\delta_2$ will decrease $R_0$ by 40.36%. The parameters $\gamma_1$ and $\delta_1$ has no influence on the basic reproductive number.

### 5.2 Numerical Simulation

We perform the numerical simulation to compare our results with the real data of Ghana and a population with an underlying condition of diabetes. A starting point of our simulation is 12 March 2020 where the authorities of Ghana confirmed the first two cases of the COVID-19 [35]. According to 2020 population and housing census, the population of Ghana is 30.8 million [36] and between 3.3 and 6% of the Ghanaian population has diabetes [14]. In agreement, in our model we consider the total population under study N=30.8 million. We choose the following initial conditions: $D(0) = 0.06N = 1,848,000$, $I(0) = 2, E(0) = 0, C(0) = 0, R(0) = 0$ and $(0) = N - D(0) - I(0) = 28,951,998$. Using matlab fourth-order Runge-Kutta method, the simulations performed are displayed in Figs. 4 – 9.

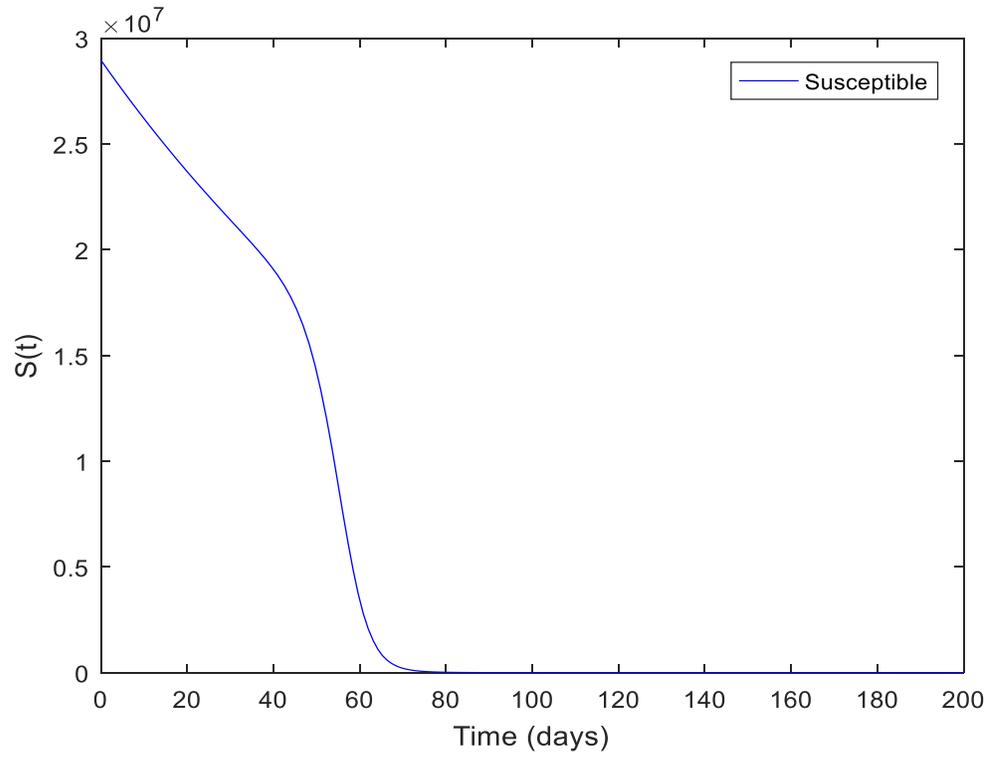

Fig. 4: Behaviour of susceptible individual

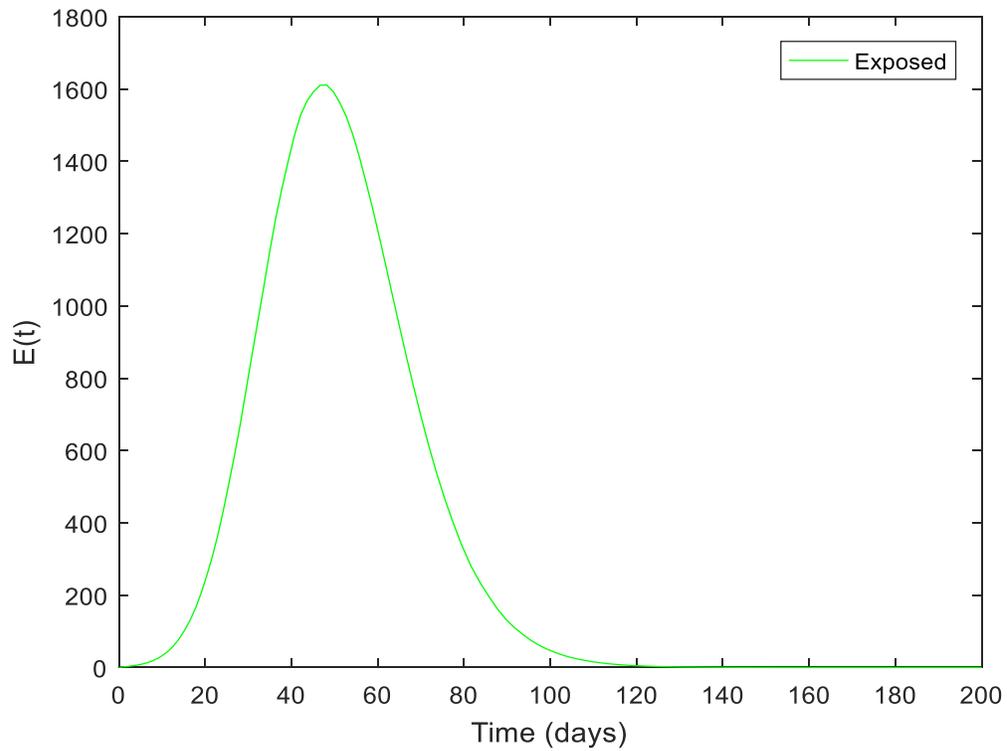

Fig. 5: Behaviour of the individual exposed to COVID-19

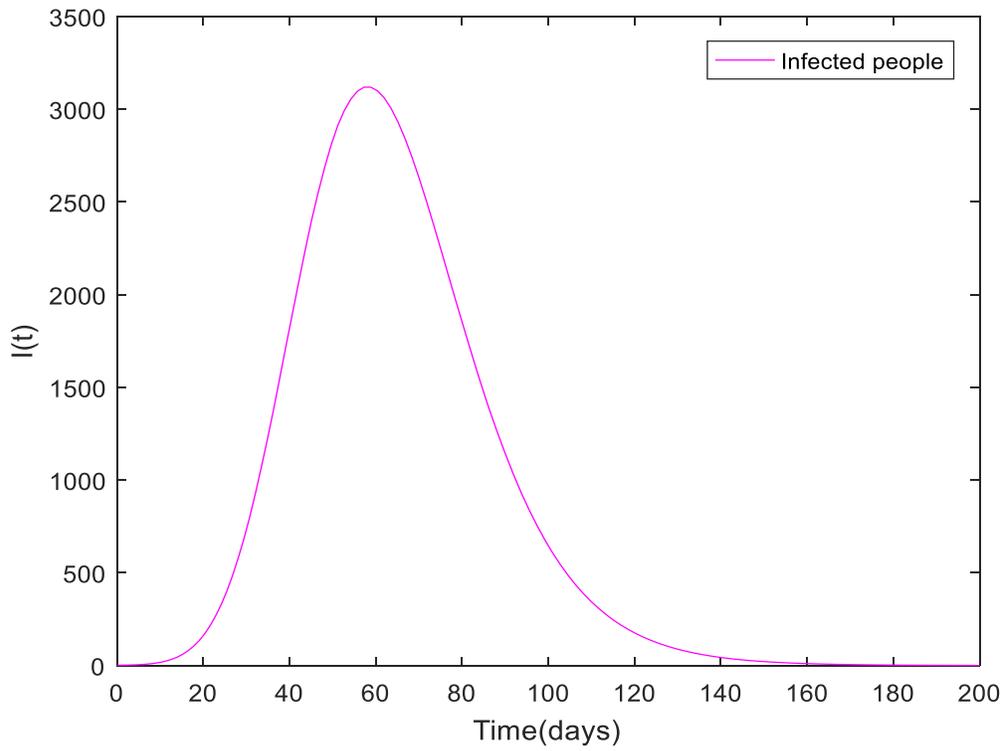

Fig. 6: Behaviour of the COVID-19 infected individuals.

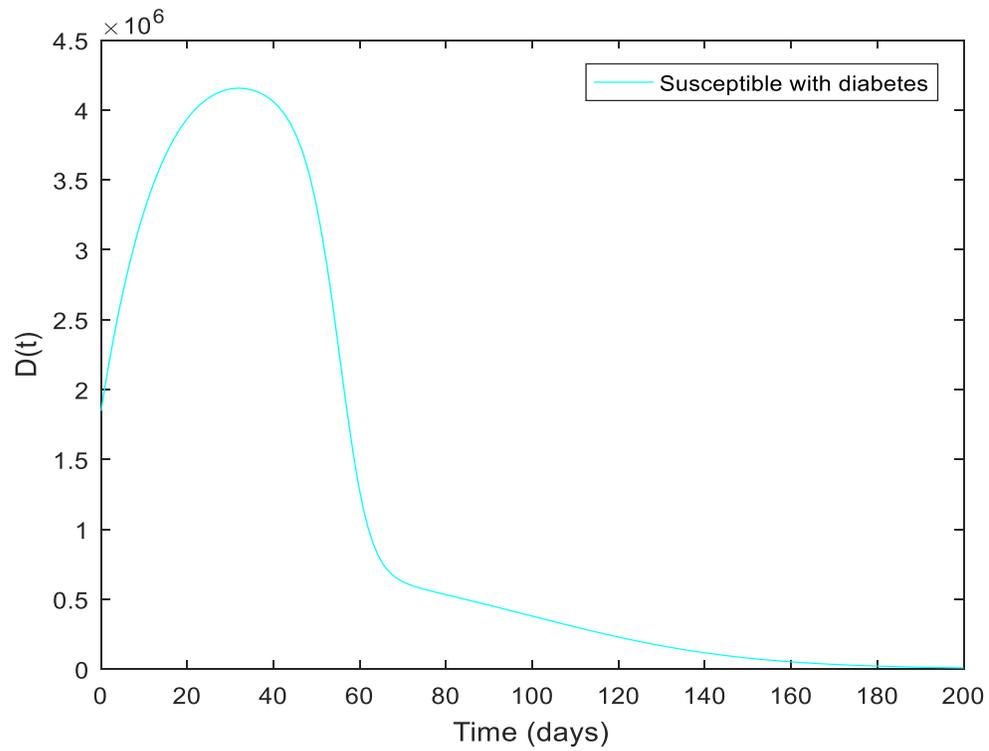

Fig. 7: Behaviour of the susceptible with diabetes

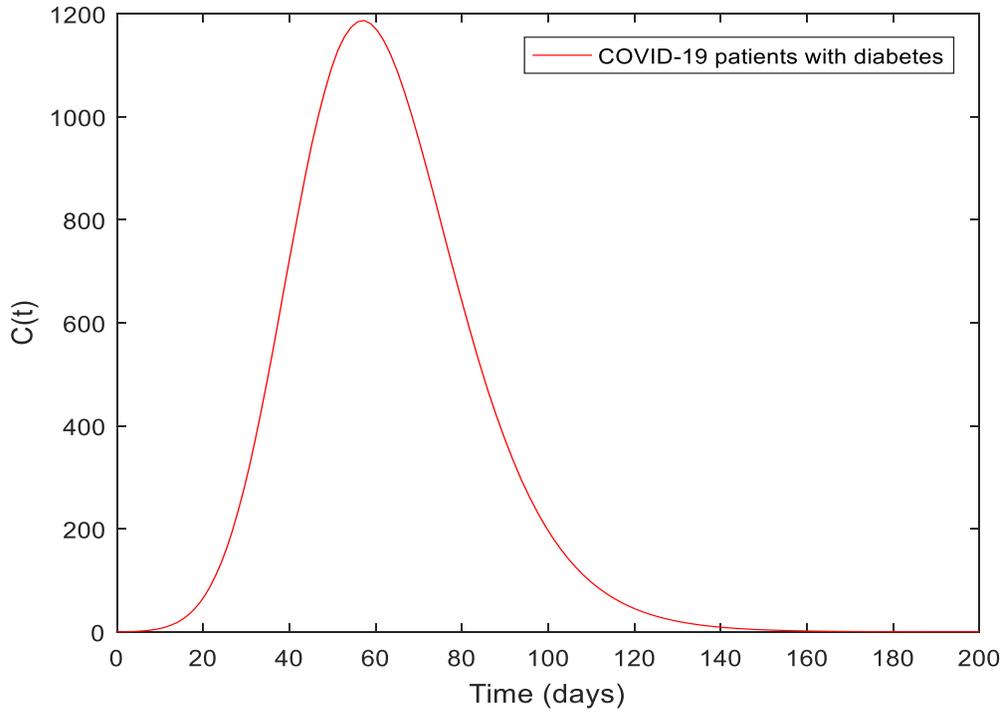

Fig. 8: Behaviour of the COVID-19 patients with diabetes

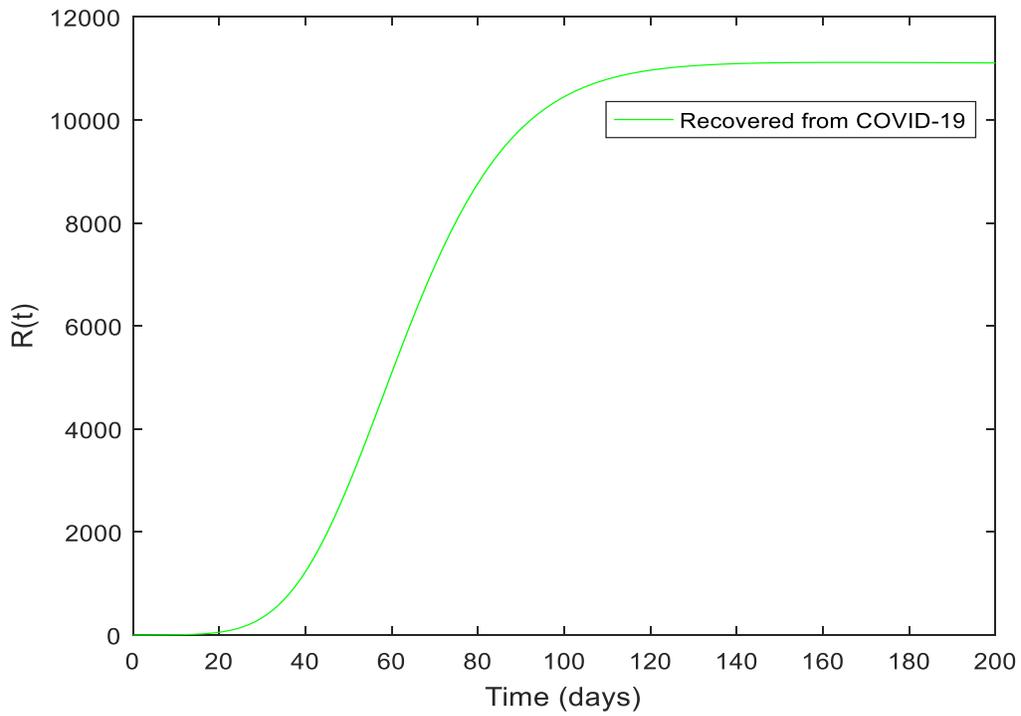

Fig. 9: Behaviour of the individuals removed from COVID-19

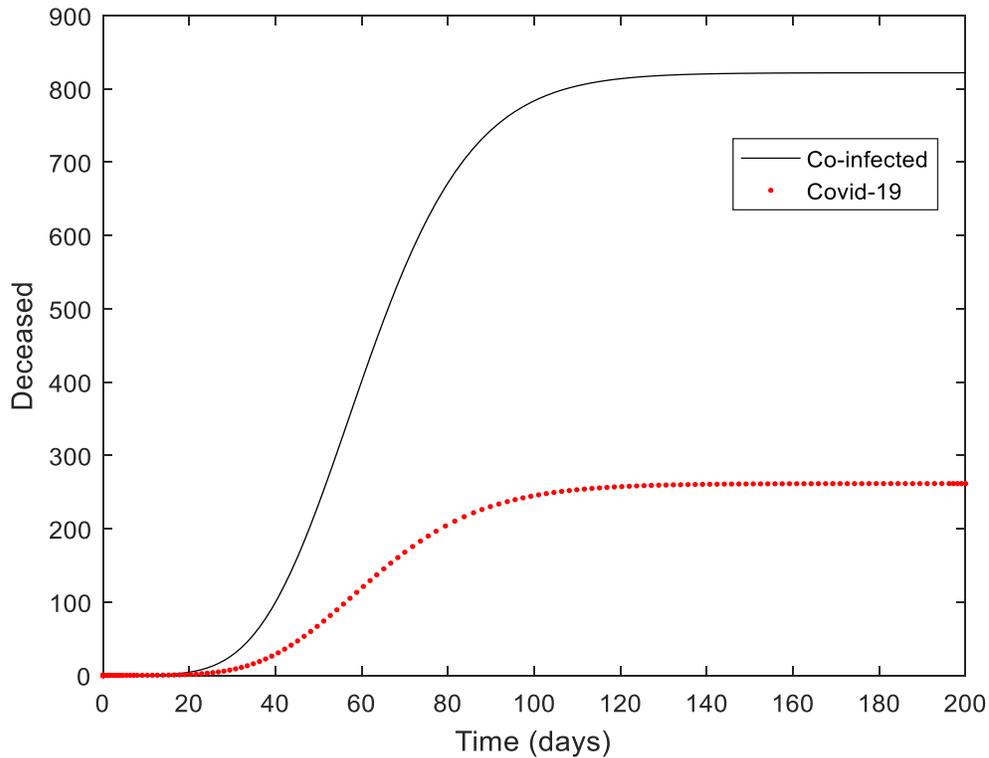

Fig. 10: Deceased COVID-19 individuals

Figure 4 depicts the behaviour of the susceptible individuals. It can be observed from Fig. 4 that the susceptible population decays with time. Figure 5 shows the behaviour of individuals exposed to COVID-19, the exposed population increases, reaches its peak in the first 50 days then declines with time. In figure 6, the population of COVID-19 infected individuals increases in the first 60 days then declines with time. Figure 7 depicts the behaviour of the susceptible with diabetes, the population of susceptible with diabetes decays and reaches zero at the end of the 200-day period. In figure 8 the population of COVID-19 patients with diabetes increases in the first 60 days, then declines with time. Figure 9 depicts the behaviour of individuals removed from COVID-19. The recovered population increases exponentially with time, reaches its maximum on the 100$^{th}$ day then levels off. Figure 10 depicts individuals deceased as a result of the COVID-19 disease. It can be seen that the probability of death in the population of COVID-19 patients with the underlying condition of diabetes is significantly higher as compared to COVID-19 infected population with no diabetes.

## 5 Optimal Control Model

In this section, we incorporate optimal controls into the system (1). We consider two preventive control measures; control ($u_1$) which represents lockdown, and control ($u_2$) which represents vaccination of susceptible individuals. To minimize the infections, countries first adopted the lockdown policy by controlling the movement of people. In this case, interactions among the susceptible individuals, susceptible with diabetes, COVID-19 infected individuals and COVID-19 patients with diabetes were reduced. To include this lockdown in the model, we replaced the parameter $\beta$ with $(1-u_1)\beta$, where $0 \leq u_1 \leq 1$. If there are no lockdown, then $u_1 = 0$ and if there are total lockdown then $u_1 = 1$. After the lockdown, countries have resulted to the vaccination of individuals to allow free movement of people and easing most restrictions. In this case we vaccinate the susceptible individuals both with an underlying condition and without an underlying condition. We include the time-dependent controls into the system (1) and we have

$$\frac{dS}{dt} = \Omega - (1-u_1)\beta\left(\frac{I+C}{N}\right)S - (\mu + \lambda)S - u_2 S,$$

$$\frac{dD}{dt} = \lambda S + \gamma_1 C - (1-u_1)\beta\left(\frac{I+C}{N}\right)D - (\delta_2 + \mu)D - u_2 D,$$

$$\frac{dE}{dt} = (1-u_1)\beta\left(\frac{I+C}{N}\right)S + (1-u_1)\beta\left(\frac{I+C}{N}\right)D - (\varphi + \mu)E, \quad (11)$$

$$\frac{dI}{dt} = \alpha\varphi E - (\delta + \mu + \gamma)I,$$

$$\frac{dC}{dt} = (1-\alpha)\varphi E - (\delta_1 + \mu)C,$$

$$\frac{dR}{dt} = \gamma I - \mu R.$$

## 5.2 Analysis of the Optimal Control Model

We analyse the behaviour of the system (11). The objective function for fixed time $t_f$ is given by

$$J(u_1, u_2) = \int_0^{t_f} [f_1 S(t) + f_2 E(t) + f_3 I(t) + f_4 C(t) + f_5 D(t) + \frac{1}{2}(T_1 u_1^2 + T_2 u_2^2)]dt, \qquad (12)$$

where $f_1, f_2, f_3, f_4$ and $f_5$ are the relative weights and $T_1$ and $T_2$ are the relative cost associated with the controls $u_1$ and $u_2$. The final time of the control is $t_f$. The aim of the control is to minimize the cost function.

$$J(u_1^*, u_2^*) = \min_{u_1, u_2 \in U} J(u_1, u_2), \qquad (13)$$

subject to the system (11), where $0 \leq (u_1, u_2) \leq 1$ and $t \in (0, t_f)$. In other to derive the necessary condition for the optimal control, Pontryagin maximum principle given in [4] was used. This principle converts system (11) - (13) into a problem of minimizing a Hamiltonian H, defined by

$$\begin{aligned}
H = &\, f_1 S(t) + f_2 D(t) + f_3 E(t) + f_4 I(t) + f_5 C(t) + \frac{1}{2}(T_1 u_1^2 + T_2 u_2^2) \\
&+ \Lambda_S \{(\Omega - (1-u_1)\beta\left(\frac{I+C}{N}\right) S - (\mu + \lambda) S - u_2 S\} \\
&+ \Lambda_D \{\lambda S + \gamma_1 C - (1-u_1)\beta\left(\frac{I+C}{N}\right) D - (\delta_2 + \mu) D - u_2 D\} \\
&+ \Lambda_E \{(1-u_1)\beta\left(\frac{I+C}{N}\right) S + (1-u_1)\beta\left(\frac{I+C}{N}\right) D - (\varphi + \mu) E\} \qquad (14) \\
&+ \Lambda_I \{\alpha \varphi E - (\delta + \mu + \gamma) I\} \\
&+ \Lambda_C \{(1-\alpha)\varphi E - (\delta_1 + \mu) C\} \\
&+ \Lambda_R \{\gamma I - \mu R\},
\end{aligned}$$

where $\Lambda_S, \Lambda_D, \Lambda_E, \Lambda_I, \Lambda_C,$ and $\Lambda_R$ represent the adjoint variables or co-state variables. The system of equations is derived by taking into account the correct partial derivatives of the system (14) concerning the associated state variables.

**Theorem 4**: Given optimal control $u_1^*, u_2^*$ and corresponding solution $S^*, D^*, E^*, I^*, C^*, R^*$ of the corresponding state system (11) – (12) that minimizes $J(u_1, u_2)$ over U, there exist adjoint variables $\Lambda_S, \Lambda_D, \Lambda_E, \Lambda_I, \Lambda_C, \Lambda_R$, satisfying

$$-\frac{d\Lambda_i}{dt} = \frac{\partial H}{\partial i}, \qquad (15)$$

where $i = \Lambda_S, \Lambda_E, \Lambda_A, \Lambda_Q, \Lambda_V, \Lambda_R$, with the transversality conditions

$$\Lambda_S(t_f) = \Lambda_D(t_f) = \Lambda_E(t_f) = \Lambda_I(t_f) = \Lambda_C(t_f) = \Lambda_R(t_f) = 0$$

*Proof:* The differential equations characterized by the adjoint variables are obtained by considering the right-hand side differentiation of the system (14) determined by the optimal control. The adjoint equations derived are given as

$$\frac{d\Lambda_S}{dt} = -f_1 + (1-u_1)\beta\left(\frac{I+C}{N}\right)[\Lambda_S - \Lambda_E] + (\mu + u_2)\Lambda_S + \lambda[\Lambda_S - \Lambda_D],$$

$$\frac{d\Lambda_D}{dt} = -f_2 + (1-u_1)\beta\left(\frac{I+C}{N}\right)[\Lambda_D - \Lambda_E] + (\delta_2 + \mu + u_2)\Lambda_D,$$

$$\frac{d\Lambda_E}{dt} = -f_3 + (\varphi + \mu)\Lambda_E + \alpha\varphi[\Lambda_C - \Lambda_I] - \varphi\Lambda_C,$$

$$\frac{d\Lambda_I}{dt} = -f_4 + (1-u_1)\frac{\beta S}{N}[\Lambda_S - \Lambda_E] + (1-u_1)\beta\frac{D}{N}[\Lambda_D - \Lambda_E] + (\gamma + \mu + \delta)\Lambda_I - \gamma\Lambda_R,$$

$$\frac{d\Lambda_C}{dt} = -f_5 + (1-u_1)\frac{\beta S}{N}[\Lambda_S - \Lambda_E] + (1-u_1)\beta\frac{D}{N}[\Lambda_D - \Lambda_E] - \gamma\Lambda_D + (\mu + \delta_1)\Lambda_C,$$

$$\frac{d\Lambda_R}{dt} = \mu\Lambda_R.$$

(16)

By obtaining the solution for $u_1^*$ and $u_2^*$ subject to the constraints, we have

$$0 = \frac{\partial H}{\partial u_1} = -T_1 u_1 + \beta\left(\frac{I+C}{N}\right)S[\Lambda_E - \Lambda_S] + \beta\left(\frac{I+C}{N}\right)D[\Lambda_E - \Lambda_D],$$

$$0 = \frac{\partial H}{\partial u_1} = -T_2 u_2 + S\Lambda_S + D\Lambda_D.$$

(17)

This gives

$$u_1^* = \min\left(1, \max\left(0, \frac{\beta\left(\frac{I+C}{N}\right)S[\Lambda_E - \Lambda_S] + \beta\left(\frac{I+C}{N}\right)D[\Lambda_E - \Lambda_D]}{T_1}\right)\right),$$

$$u_2^* = \min\left(1, \max\left(\frac{S\Lambda_S + D\Lambda_D]}{T_2}\right)\right).$$

(18)

Hence the theorem is proved.

## 6 Numerical Analysis of the Optimal Control Model

In this section, we analysed numerically the behaviour of the optimal control model (11) using the method of forward-backward sweep method as in [43]. We develop a numerical scheme that uses matlab fourth order Runge-Kutta method [40 - 43] to solve the model's optimality system.

### 6.1 Optimal Control $(u_1)$ of the Model

The control $u_1$ which represents lockdown is optimized throughout the period of 200 days whilst setting $u_2 = 0$. Using the parameter values given in Table 1 and same initial conditions, the simulation performed are displayed in Figs. 11 – 16 which depicts the dynamical behaviour of all the compartments.

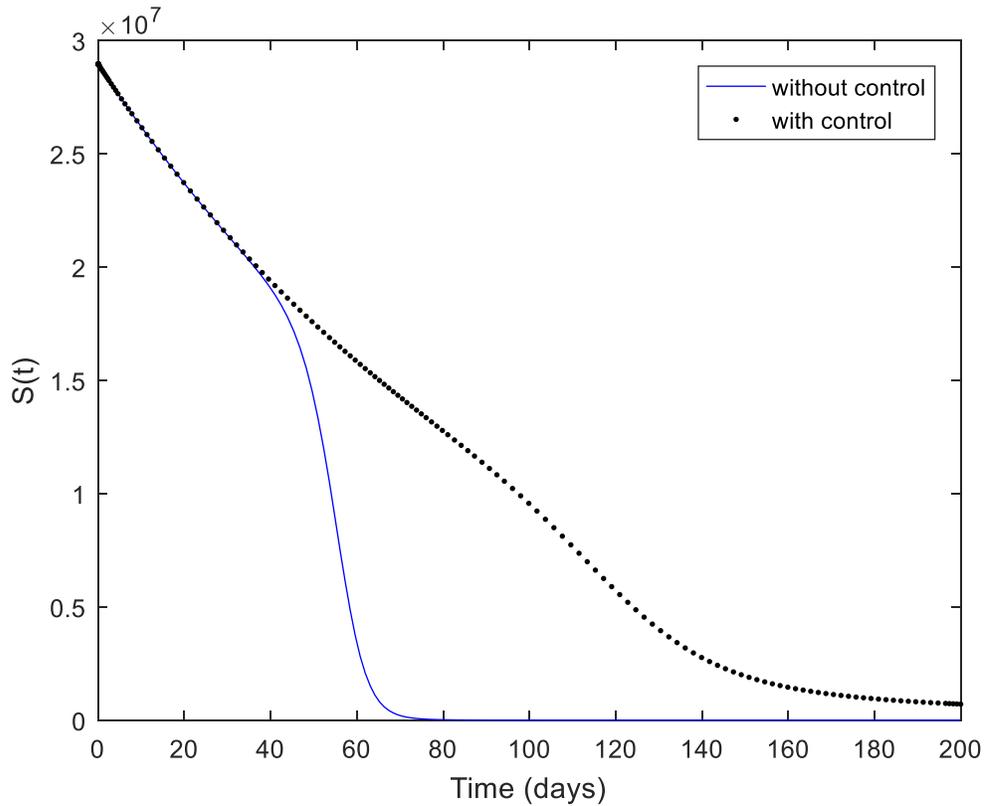

Fig. 11: Behaviour of the susceptible population with and without lockdown control

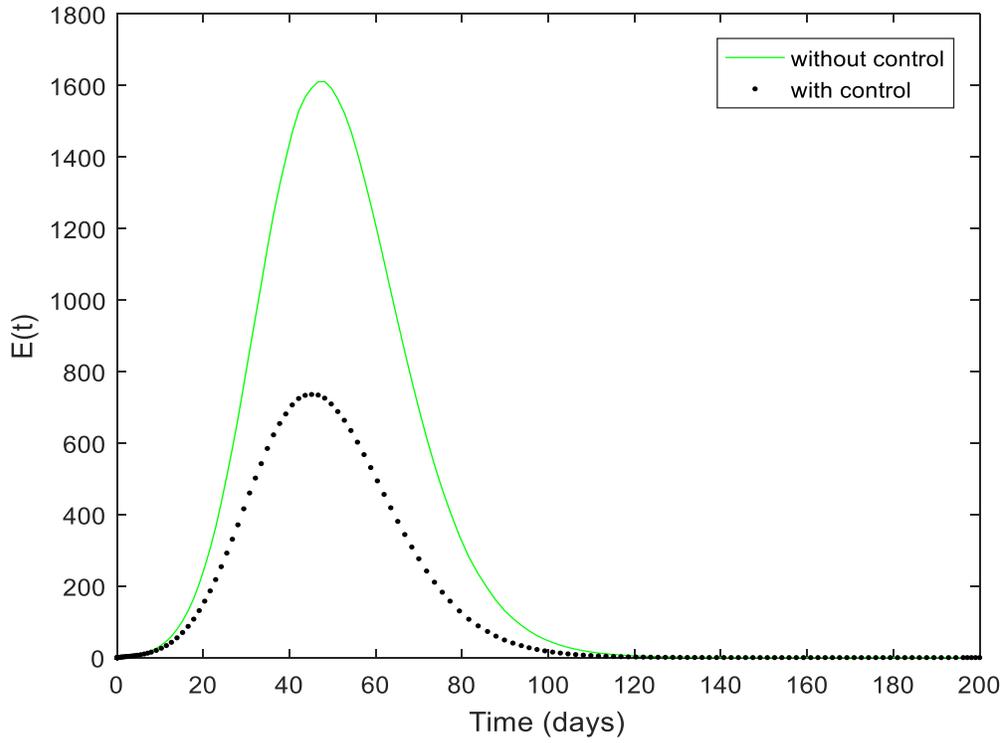

Fig. 12: Behaviour of the exposed individuals with and without lockdown control

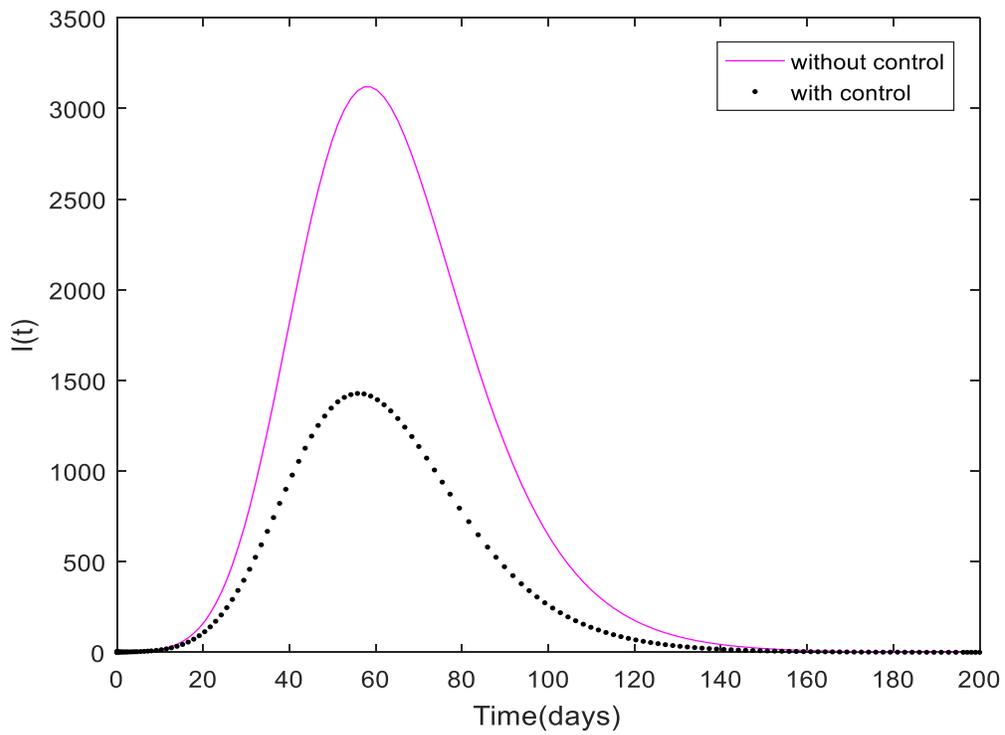

Fig. 13: Behaviour of COVID-19 infected individuals with and without lockdown control

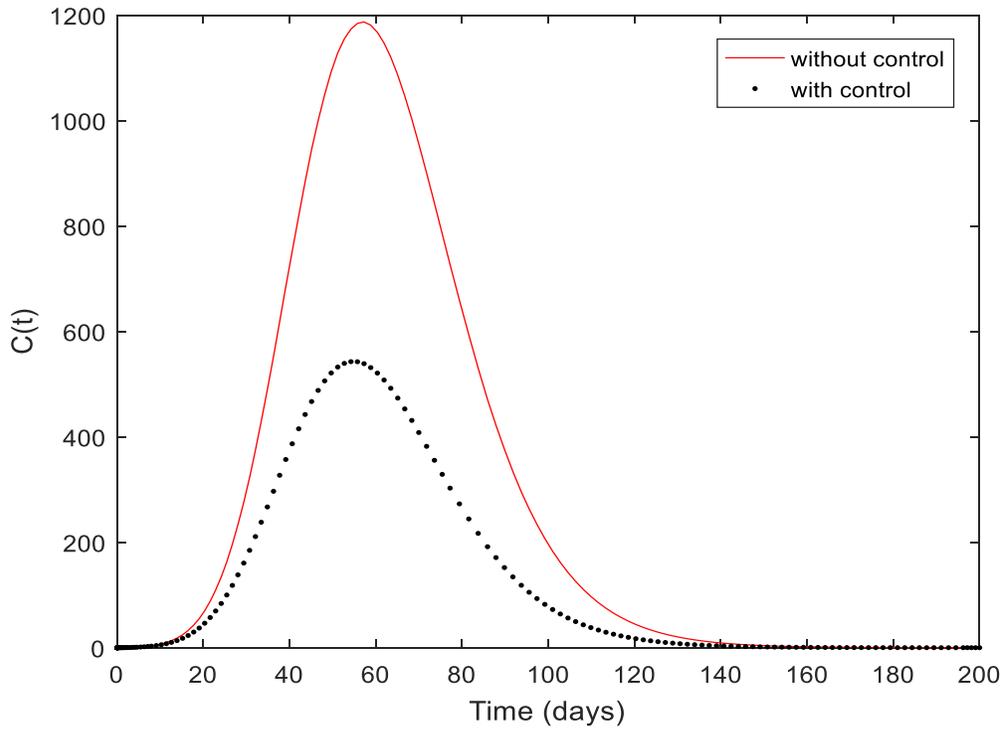

Fig. 14: Behaviour of COVID-19 patients with diabetes with and without lockdown control

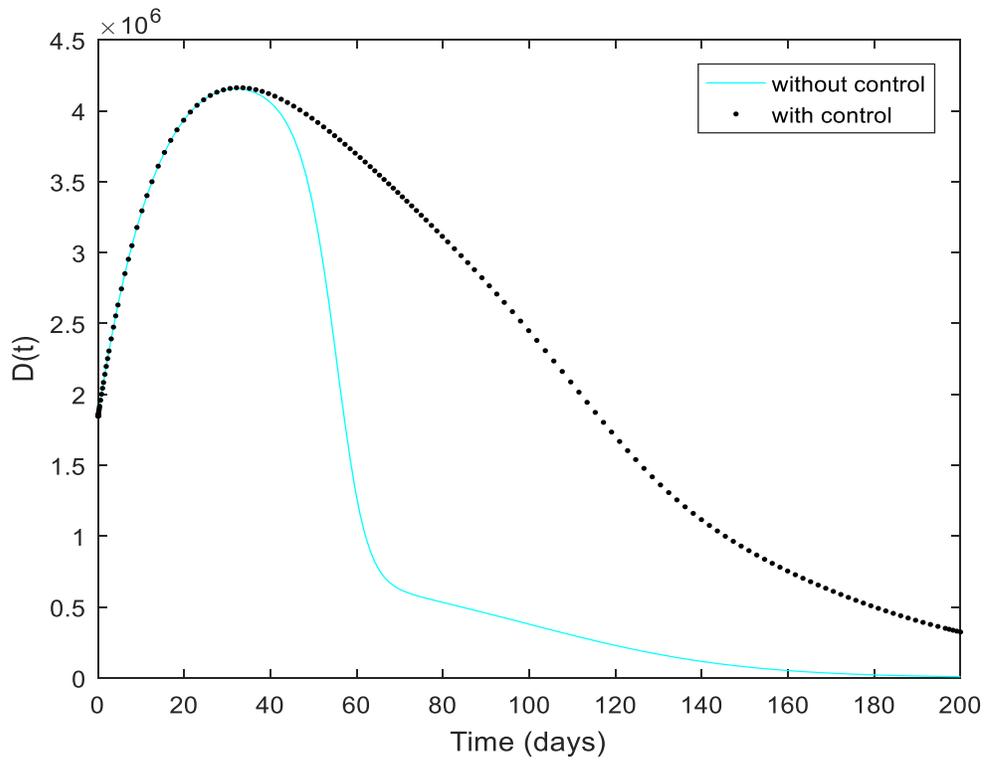

Fig. 15: Behaviour of susceptible individuals with diabetes with and without lockdown control

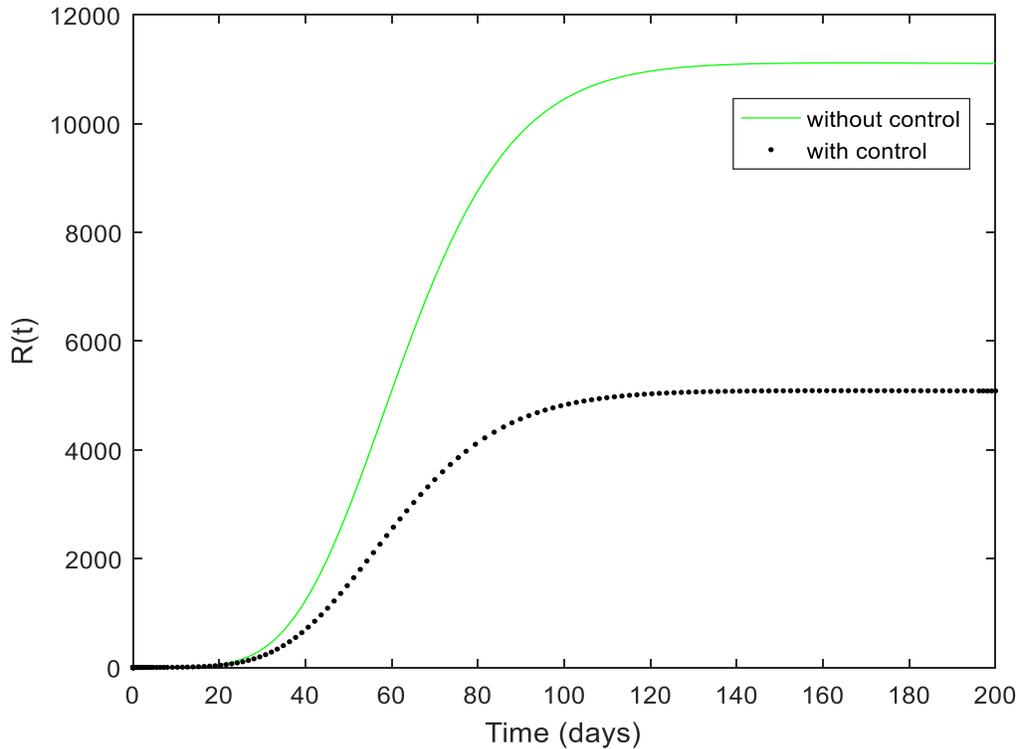

Fig. 16: Behaviour of the individuals removed from COVID-19 with and without lockdown control

In figure 11, it can be observed that the lockdown sustains the susceptible population for the entire period of 200 days as compared to no lockdown situation which decays the susceptible population by the end of the 70$^{th}$ day. In figure 12, the behaviour of the COVID-19 exposed individuals with and without the imposition of a lockdown is depicted. it can be observed that the imposition of a lockdown on the population, reduces the number of individuals that get exposed to the COVID-19 from the peak of 1600 to less than 800 on the 40$^{th}$ day. Figure 13 depicts the behaviour of COVID-19 infected individuals with and without the lockdown. The lockdown reduces the number of COVID-19 infected individuals from the peak of 3200 to 1500 on the 60$^{th}$ day. Figure 14 depicts the behaviour of COVID-19 patients with diabetes with and without the lockdown. It can be observed from figure 12, the lockdown control reduces the number of people with the diabetes who contract COVID-19 with the peak reaching 500 on the $60^{th}$ day as compared with 1200 in a situation without the lockdown. Figure 15 depicts the behaviour of the susceptible with diabetes with and without the lockdown control. The lockdown control sustains the population of individuals with the diabetes as it decays within the first 160 days in a situation where there is no lockdown. Figure 16 depicts the behaviour of individuals removed from COVID-19 with and without the lockdown control. There is a decline in the population of individuals removed from COVID-19 when the lockdown control is implemented. This is so because with the lockdown movement of susceptible people are restricted and hence less infections which translate to the decline in the recoveries.

## 6.2  Control $(u_2)$ of COVID-19

We now focus our attention on vaccination control $(u_2)$. The goal is to reduce the number of people who contract COVID-19. Using the same initial conditions and parameter values given in Table 1, the results of the simulations are displayed in Figs. 17 – 22.

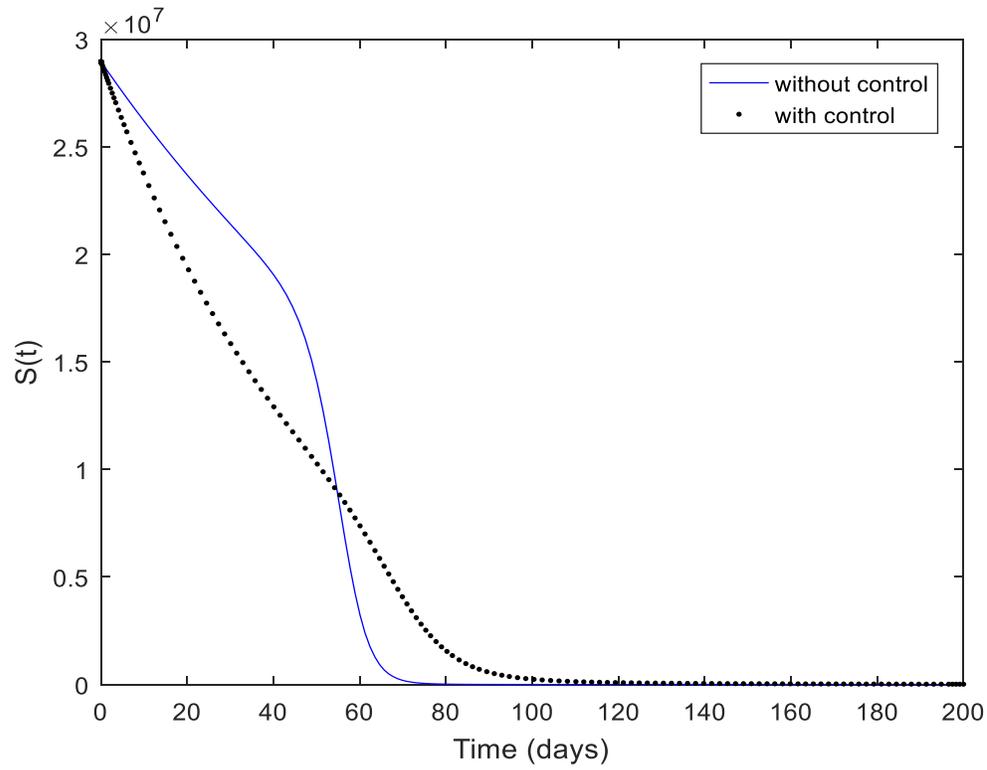

Fig. 17: Behaviour of the susceptible population with and without vaccination control

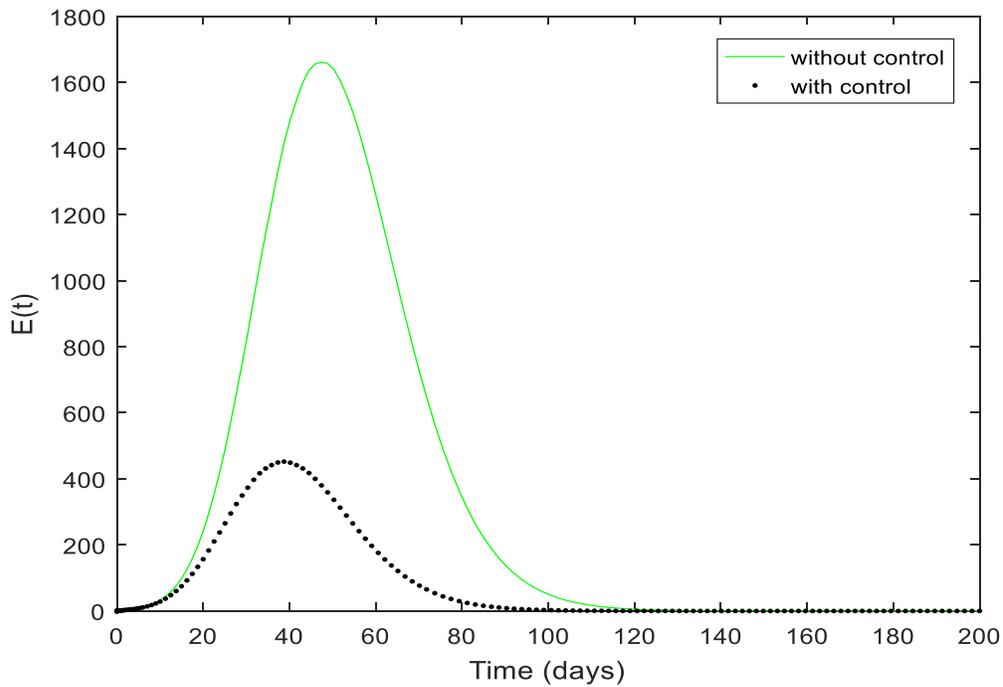

Fig. 18: Behaviour of the individuals exposed to COVID-19 with and without vaccination control

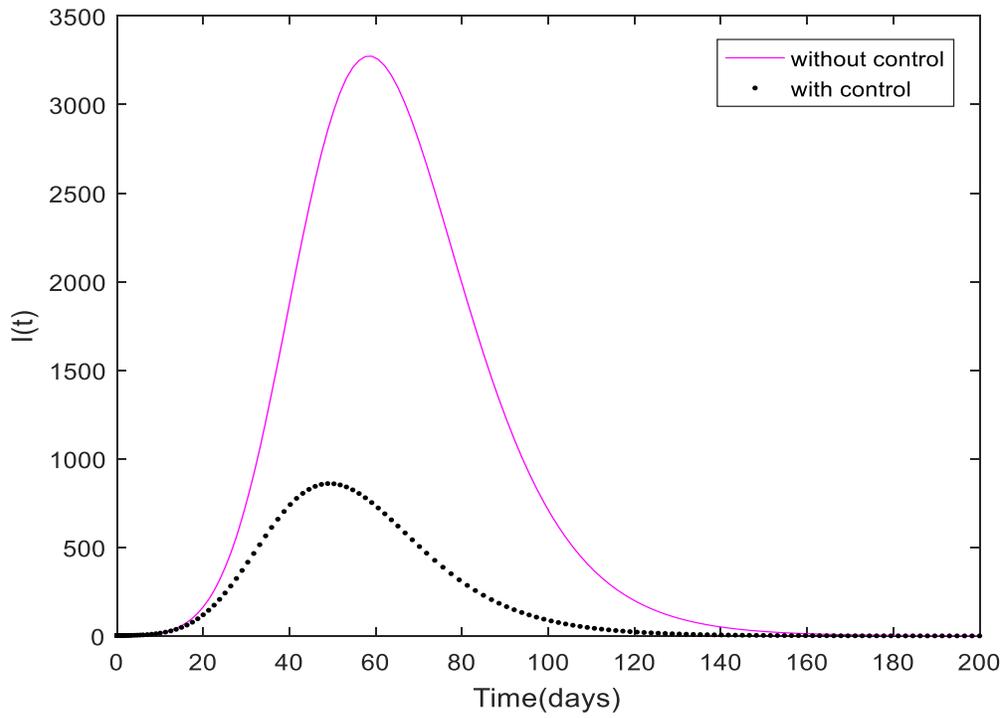

Fig. 19: Behaviour of the COVID-19 infected individuals with and without vaccination control

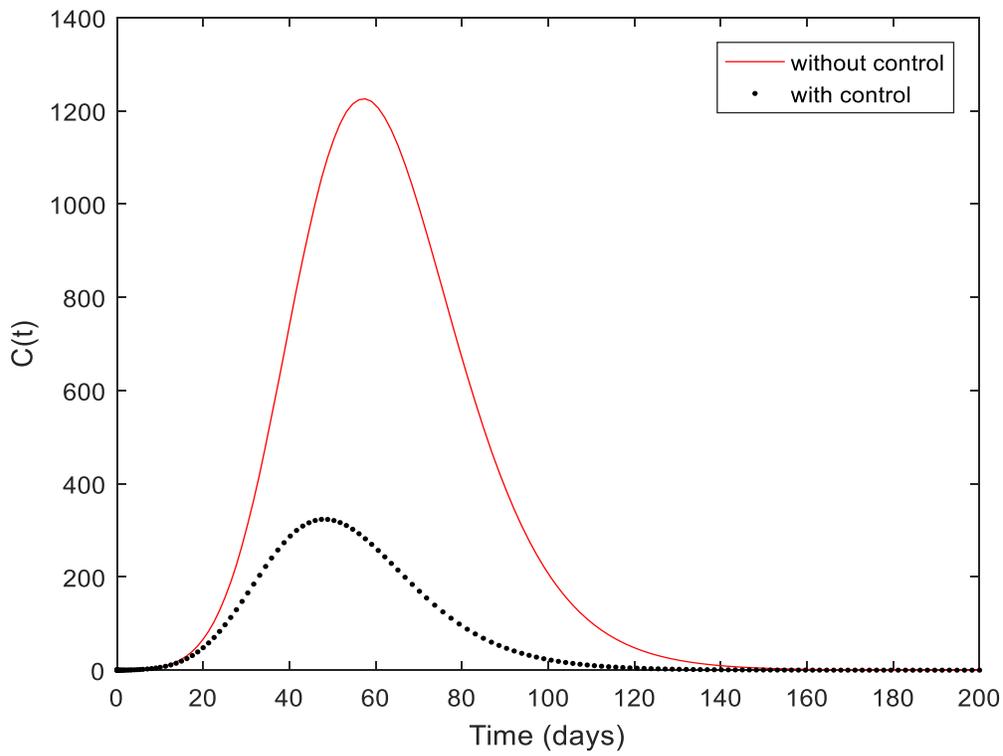

Fig. 20: Behaviour of COVID-19 patients with diabetes with and without control

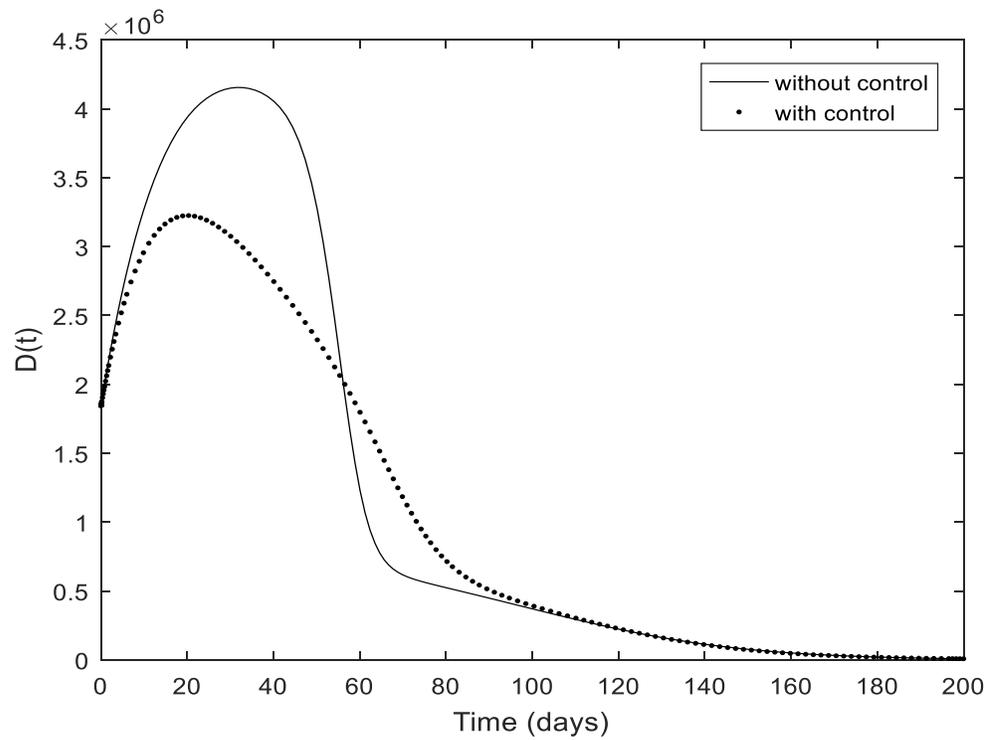

Fig. 21: Behaviour of susceptible individuals with diabetes, with and without control

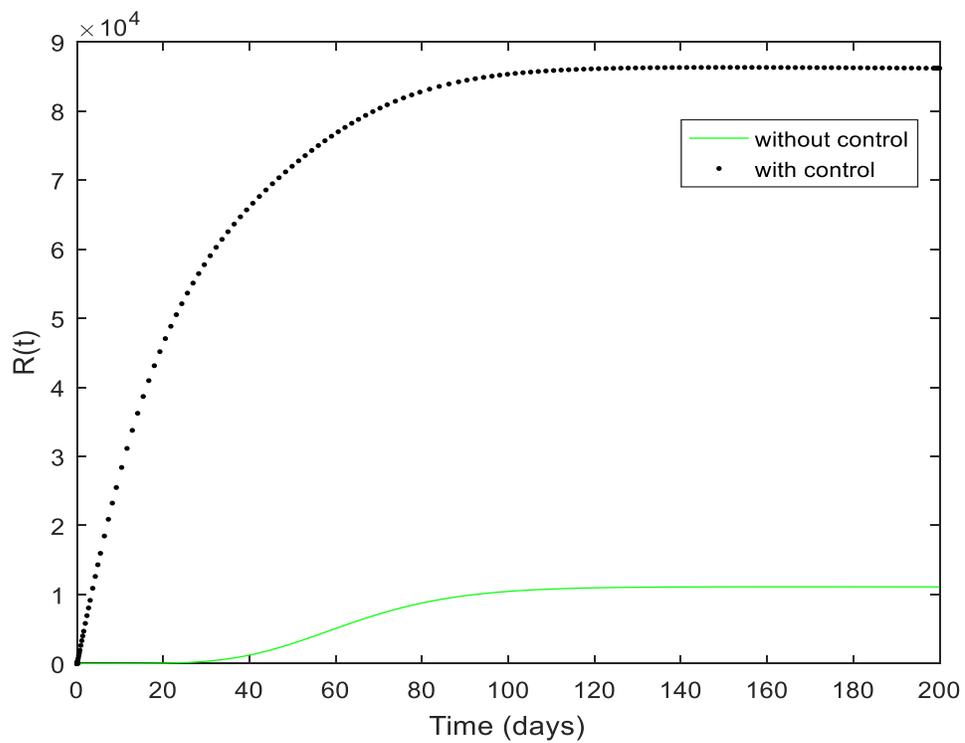

Fig. 22: Behaviour of the individuals removed from COVID-19 with and without vaccination control

Figure 17 depicts the dynamic behaviour of the susceptible individuals when there is a vaccination control within 200 days. The vaccination reduces the number of individuals who become susceptible to COVID-19 and thereby developing immunity to the disease. Figure 18 depicts the behaviour of individuals exposed to COVID-1 when there is a vaccination control within 200 days. It can be observed that, the number of individuals who become exposed to COVID-19 declines as a result of individuals getting vaccinated. The number reduces from the peak of 1600 to 400 and then decays at the end of the $100^{th}$ day. Figure 19 depicts the behaviour of COVID-19 infected individuals when there is a vaccination control within 200 days. The number of individuals who contract COVID-19 reduces drastically as vaccination control is implemented. Figure 20 depicts the behaviour of COVID-19 patients with diabetes with and without vaccination control. With the vaccination of individuals with an underlying condition of diabetes, it can be seen from figure 20 that there is a decline in the number of individuals who contract COVID-19 as compared to a situation where there is no vaccination. The population of COVID-19 patients with diabetes declines at the peak of 1200 to a peak of 300 on the $50^{th}$ day. Figure 21 depicts the behaviour of susceptible with diabetes when vaccination control is implemented. There is a decline in the number individuals with diabetes who become susceptible to the COVID-19 within the first 60 days as they become immune after the vaccination. Figure 22 depicts the behaviour of individuals removed from COVID-19. The number of individual removed from COVID-19 increases as those vaccinated joins the removed class.

## 7  Conclusion

In this paper, we have formulated and analysed a mathematical model which describes the transmission of COVID-19 in a population with an underlying condition of diabetes. The basic properties of the model were explored. The positivity and boundedness of the solution, the basic reproductive number, equilibrium points and stability of the equilibrium points and sensitivity analysis of the model were examined. The basic reproduction number of the model was found to be $R_0 = 1.4722$. Results of the numerical simulation suggest a greater number of individuals deceased when the infected individual had an underlying condition of diabetes. Time-dependent controls were incorporated into the model with the sole aim of determining the best strategy in curbing the spread of the disease. Pontrygin's maximum principle was used to characterize vital conditions of the optimal control model. The numerical simulation of the optimal control model suggests the lockdown and vaccination when implemented reduces COVID-19 infection in the population. The lockdown control minimizes the rate of decay of the susceptible individuals whereas the vaccination led to a number of susceptible individuals becoming immune to infections. In all the two preventive control measures were both effective in curbing the spread of the disease. Future research should include fractional-order derivative. Fractional models provide a more accurate and deeper analysis of the model's behaviour than a classical integer model [27]. Also more attention should be paid to individuals with the underlying condition of diabetes as the number of deaths was significantly high.

**Data Availability**

The data/information supporting the formulation of the mathematical model in this paper are/is from Ghana health service website: https://www.ghs.gov.gh/covid19/ which has been cited in the manuscript.

**Declaration of interest**

There are no competing interest

**Funding**

None

## Acknowledgement

This Manuscript was submitted as a pre-print in the link https://arxiv.org/ftp/arxiv/papers/2201/2201.08224.pdf and has been referenced [28].